\documentclass[sigconf, authorversion]{acmart}
\AtBeginDocument{%
  \providecommand\BibTeX{{%
    \normalfont B\kern-0.5em{\scshape i\kern-0.25em b}\kern-0.8em\TeX}}}

\acmConference[ICAIL]{19th International Conference on Artificial
Intelligence and Law}{June 19--23,
  2023}{Braga, Portugal}
\copyrightyear{2023} 
\acmYear{2023} 
\setcopyright{acmlicensed}\acmConference[ICAIL 2023]{Nineteenth International Conference on Artificial Intelligence and Law}{June 19--23, 2023}{Braga, Portugal}
\acmBooktitle{Nineteenth International Conference on Artificial Intelligence and Law (ICAIL 2023), June 19--23, 2023, Braga, Portugal}
\acmPrice{15.00}
\acmDOI{10.1145/3594536.3595166}
\acmISBN{979-8-4007-0197-9/23/06}

\begin{document}

\title{JusticeBot: A Methodology for Building Augmented Intelligence Tools for Laypeople to Increase Access to Justice}



\author{Hannes Westermann}
\orcid{0000-0002-4527-7316}
\affiliation{%
  \institution{Cyberjustice Laboratory\\ Faculté de droit \\ Université de Montréal}
  \city{Montréal}
  \country{Canada}
}
\email{hannes.westermann@umontreal.ca}

\author{Karim Benyekhlef}
\orcid{0000-0001-9390-556X}
\affiliation{%
  \institution{Cyberjustice Laboratory\\ Faculté de droit\\ Université de Montréal}
  \city{Montréal}
  \country{Canada}
}


\begin{abstract}
  Laypeople (i.e. individuals without legal training) may often have trouble resolving their legal problems. In this work, we present the JusticeBot methodology. This methodology can be used to build legal decision support tools, that support laypeople in exploring their legal rights in certain situations, using a hybrid case-based and rule-based reasoning approach. The system ask the user questions regarding their situation and provides them with legal information, references to previous similar cases and possible next steps. This information could potentially help the user resolve their issue, e.g. by settling their case or enforcing their rights in court. We present the methodology for building such tools, which consists of discovering typically applied legal rules from legislation and case law, and encoding previous cases to support the user. We also present an interface to build tools using this methodology and a case study of the first deployed JusticeBot version, focused on landlord-tenant disputes, which has been used by thousands of individuals.
\end{abstract}




\maketitle

\section{Introduction}
\label{intro}
Most people in society will face a legal issue at some point in their lives. Prevalent such issues include consumer, debt, employment and family problems \cite{currie2009legal, farrow2016everyday}. These issues are not trivial - in a study conducted in Canada in 2009, almost 60\% of individuals facing a legal issue indicated that it made their life at least somewhat difficult \cite{currie2009legal}.

However, individuals often seem to have trouble resolving these issues. The study conducted in 2009 showed that many individuals did nothing to address their issue, frequently believing that nothing could be done or not knowing how to get help. Other individuals tried to resolve their issues on their own, but often lamented the lack of assistance and public legal information \cite{currie2009legal}. Only a small minority of the people with legal issues used the court system to solve them. A study conducted in 2021 showed that only 21\% of problems that occurred in the three years before the study had been resolved \cite{savage2022experiences}. This is a global problem - the UN Task Force for Justice indicates that around 1.5 billion individuals have justice problems that they cannot resolve \cite{noauthor_justice_2019}. 

Individuals that do try to use the court system to resolve their issues often face a procedure that is overwhelming and frustrating for them. Semple describes three costs of seeking civil justice in Canada. The financial cost can be substantial, even eclipsing the value of the dispute, which leads many individuals to self-represent. This, in turn, can exacerbate the temporal and emotional cost of going to court, as people spend a significant amount of time on the judicial process, and are faced with an unusual, stressful and frustrating experience \cite{semple2015cost}. Self-represented litigants, a rapidly growing group, are often confused by which forms to fill out \cite{macfarlane2013national}, or which facts they need to prove \cite{branting2020judges}. Pro-se litigants also cause significant strain on the courts \cite{zeleznikow2016can}.

Unresolved legal issues can cause issues for the individual and society. Studies have shown that 79\% of individuals facing a legal problem report an adverse health impact \cite{savage2022experiences}. In 2014, unresolved legal issues were estimated to cost society 746m dollars in Canada alone \cite{farrow2016everyday}.

In this work, we present the JusticeBot methodology, which seeks to improve this situation. The methodology can be used to build legal decision support tools, that give the layperson user the ability to explore their rights relating to a legal problem. The user starts by selecting a situation that has occurred (such as ``There is a bedbug infestation in my apartment'') or a goal that they want to achieve (such as ``I would like to sublet my apartment''). They are then taken through a number of questions. Finally, they are given information regarding their legal rights, an overview of the outcome of previous similar cases, and possible next steps that they can pursue to resolve their issue. However, the system never aims to predict the case of the user, or tell them what to do --- rather, it aims to augment their intelligence. The provided information can be used by the layperson to resolve their issue, for example by using it to amicably settle their case, having better information when self-representing in court, or deciding to hire a lawyer to enforce their rights. Thus, hopefully, people can gain a better understanding of the laws, leading to an increased feeling of belonging in society, and to more legal problems being resolved, which could increase trust in legal institutions.

This paper gives an overview of the methodology and the first deployed tool based on the methodology, focused on landlord-tenant disputes.


\section{Prior Work}
\label{prior_work}

\subsection{Artificial Intelligence to Increase Access to Justice}
This paper describes a methodology to build an AI system to increase access to justice. There have been a number of researchers aiming to tackle this task. For example, Branting presented the concept of advisory systems to help pro-se litigants, which was implemented in the area of protection orders \cite{branting2001advisory}. Thompson described the Justice Pathway Expert System (JPES), a concept which uses expert systems to understand the situation of the user and guide them toward avenues for resolving their issues \cite{thompson2015creating}. Zeleznikow presented the GetAid system, which aims to help lawyers determine whether an individual is eligible for legal aid \cite{zeleznikow2002using}.

The Rechtwijzer platform is an implemented system in this domain. It asks the user for their situation, and guides them towards appropriate resources, in the domains of consumer disputes and divorce \cite{bickel2015online}. The Loge-expert project aimed to support layperson users in the domain of housing law in Quebec, by encoding reasoning steps conducted by practitioners \cite{paquin1991loge}.

This work contributes to this line of research by presenting a methodology aiming to augment the intelligence of the user and a toolchain to implement such systems in  different legal areas.

\subsection{Rule-based systems}
In this work, we present a methodology to encode legal rules and reasoning steps in the form of a schema. This research is connected to a number of important prior works.

Waterman \& Peterson developed a system to provide settlement information regarding cases of product liability \cite{waterman1981models}. Allen \& Engholm presented a way to represent statutes in a digital way, using propositional logic, thereby rendering the laws easier to understand and more syntactically clear \cite{allen1977normalized}.

Sergot et al implemented the British Nationality Act in the Prolog programming language, resulting in 150 rules \cite{sergot1986british}. Walker suggested modelling legislation in the form of an implication tree, stemming from the default logic paradigm of reasoning \cite{walker2006default}. This schema was used to represent cases in Board of Veterans appeal cases \cite{walker2017semantic} and landlord-tenant disputes \cite{westermann2019using}, among others.

Satoh et al implemented PROLEG, a digitalization of the Japenese JUF-theory. The system represents possible arguments by both parties in a rule-based manner. Certain concepts are only assessed if they are raised by a party \cite{satoh2010proleg}. 

In this work, we contribute to this line of research by presenting a system that aims to discover the syntactic structure of rules in a legal area from case law, which could overcome syntactic ambiguity of rules \cite[p.~45]{ashley2017artificial}. Further, we show a way to integrate case law summaries into the rules to help the user assess the open-textured legal terms that apply to them (compare \cite{waterman1981models,sergot1986british,paquin1991loge}). 



\subsection{Case-based systems}
A key component of the JusticeBot methodology is case law, the use of which other researchers have previously explored.

Ashley developed HYPO, which used factors (``commonly observed collections of facts that tend to strengthen or weaken a plaintiff’s argument in favor of a legal claim''), captured as dimensions, to generate arguments in the field of trade secret law \cite{ashley1991reasoning}. Aleven presented CATO, which uses a simpler factor representation and adds a hierarchy to the system, linking factors to legal issues, in order to teach law students argumentation \cite{aleven1997teaching}. IBP (Issue-based prediction) introduced a domain model, which is used in conjunction with cases to predict case outcomes \cite{ashley2009automatically}. VJAP linked factors to values, making it able to reason using policy balances of competing values, and was able to generate an overall reasoning structure with any possible argument \cite{grabmair2017predicting}. \cite{rissland1991cabaret} combined rule-based reasoning with case-based reasoning in a hybrid system, able to switch between the two. \cite{al2016methodology} used abstract dialectic frameworks to represent rules extracted from cases, in order to predict new cases and generate explanations.

Other approaches have used machine learning to build models of judicial reasoning. In \cite{alarie2016using}, the researchers built a model to predict whether individuals would be seen as workers or independent contractors, achieving a high accuracy. Yin et al used a dataset of annotated decisions to predict outcomes of cases \cite{yin2020determining}. Branting et al used a projection method to build a dataset of 16k annotated datasets, which could then be used to predict the outcome of the cases in an explainable manner \cite{branting2021scalable}. In \cite{westermann2019using}, we annotated decisions in the domain of landlord-tenant disputes with factual occurrences in the cases, which were then used to predict outcomes and discover trends. We found that the machine learning model had difficulties in predicting new cases based on the annotated factors, and discussed reasons why this may be the case.

Case-based reasoning systems have also been used to predict the discretionary outcome of cases. Stranieri et al built a system to predict the distribution of property between different parties in a divorce, by combining neural networks with rule-based reasoning \cite{stranieri1999hybrid}. Dahan et al built a system to predict notice periods for individuals after a firing using machine learning \cite{dahan2020predicting}. 

The methodology presented here uses case law for three purposes. It is used to discover a schema of legal reasoning steps applied by judges. This structure is exposed to the user in an interface, to allow them to assess their own situation. Cases are also indexed with this schema, to support the user in understanding how legal criteria are applied, and give examples of previous case outcomes to the user.


\section{Using a JusticeBot tool}
\label{user_flow}
In order to set the stage for the JusticeBot methodology, we will demonstrate a possible pathway taken by the user through a JusticeBot tool. We will use the example of the JusticeBot focused on landlord-tenant disputes in Quebec, which is the first deployed version of the JusticeBot (hereafter JusticeBot TAL, since it focuses on cases within the jurisdiction of the tribunal administratif du logement, the Québec housing tribunal). 

\begin{figure}[ht]
\begin{minipage}[b]{0.45\linewidth}
\centering
\includegraphics[width=4cm]{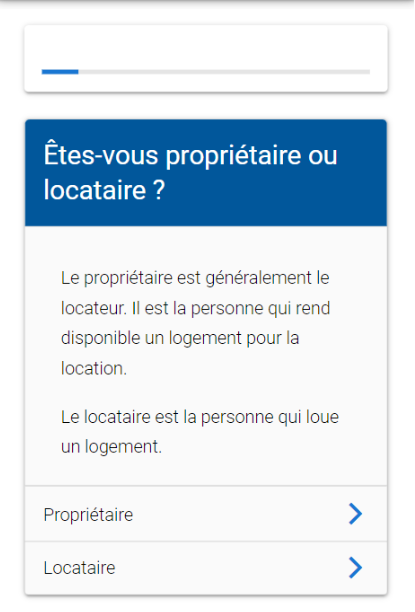}
\caption{The user is asked whether they are a landlord or a tenant.}
\label{fig:landlord_tenant}
\end{minipage}
\hspace{0.2cm}
\begin{minipage}[b]{0.45\linewidth}
\centering
\includegraphics[width=4cm]{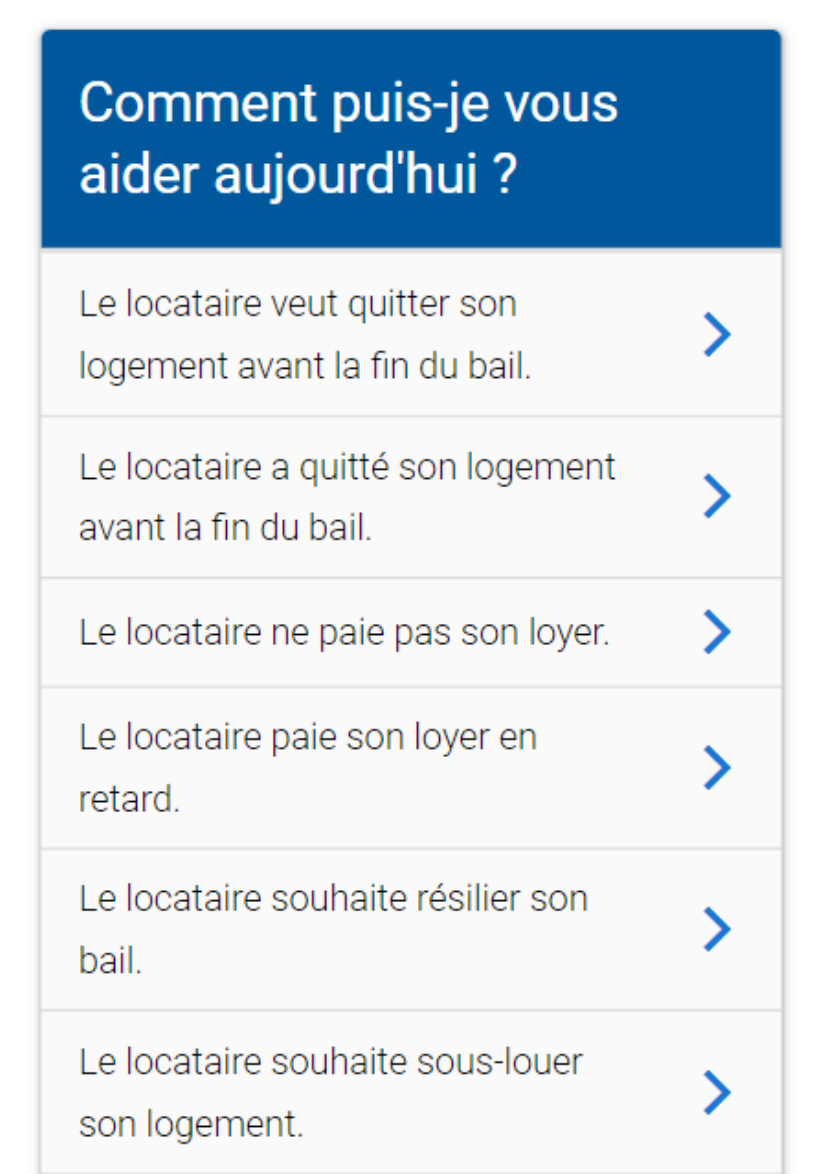}
\caption{Possible pathways for landlords.}
\label{fig:landlord_pathways}
\end{minipage}
\end{figure}

After visiting \url{https://justicebot.ca}, the user initially sees an introduction screen, that explains the scope and functioning of the system, and clarifies that it does not provide legal advice to the user. Upon clicking ``I understand'', the user is taken to the next page, which triages the situation of the user, by asking them whether they are a landlord or a tenant (Figure \ref{fig:landlord_tenant}). They are also given an explanation of what these terms mean, to help them in their assessment. For the sake of the demonstration, let us assume that the user is a landlord, who wants to learn how to deal with their tenant often being late with paying their rent. 

After selecting ``landlord'', the user will be shown a list of legal issues that the JusticeBot system can treat (figure \ref{fig:landlord_pathways}). Some of these are focused on factual situations that have occurred, such as ``My tenant has stopped paying their rent'', allowing the user to explore their rights in this situation. Other issues are described as outcomes that the user wants to achieve, such as (for tenants) ``I would like to sublet my apartment''. In our case, the user would select ``My tenant does not pay their rent''.

\begin{figure}[ht]
\begin{minipage}[b]{0.45\linewidth}
\centering
\includegraphics[width=4cm]{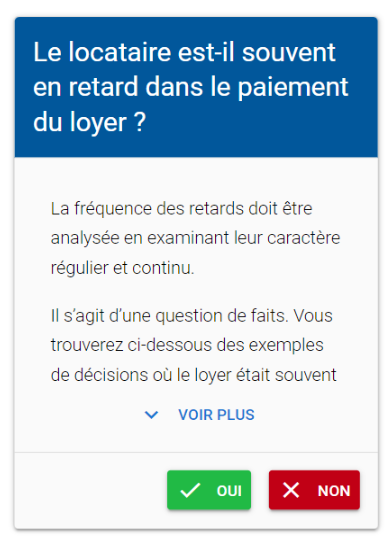}
\caption{Question asking whether the tenant is frequently late with paying their rent.}
\label{fig:freq_late}
\end{minipage}
\hspace{0.2cm}
\begin{minipage}[b]{0.45\linewidth}
\centering
\includegraphics[width=4cm]{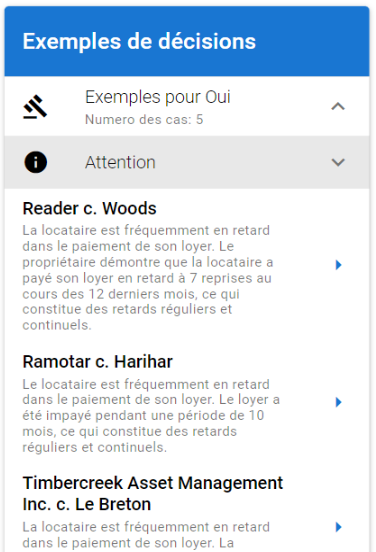}
\caption{Cases showing examples of ``frequent lateness'' being applied in previous cases.}
\label{fig:freq_late_cases}
\end{minipage}
\end{figure}

Now, the system switches to ask more in-depth questions, focused on understanding the specific situation of the user. First, they are asked whether the tenant is \textit{currently} late with paying their rent. Let us assume that the user answers ``No'' to this question. Next, they are asked whether the tenant is ``frequently'' late with paying the rent (figure \ref{fig:freq_late}). It should be noted that ``frequently'' late is in this case an open-textured legal term - the legislation does not specify what exactly is meant by it. Therefore, the user is provided with case law summaries to illustrate how this criterion was previously applied (figure \ref{fig:freq_late_cases}). For example, the user can read that the judge in a previous case found that the tenant being late 7 times in 12 months was found to be ``frequently late'', while only being late 2 times in 3 months was not found to be ``frequently late''. The user can also click through to read the entire cases. Based on these summaries, the user can assess whether their situation may warrant the criterion of frequent lateness to apply, and answer the question accordingly.

For our demonstration, let us assume that the user found that their tenant is ``frequently late'', and also answers affirmatively to the next question, that the tenant caused them a ``serious prejudice'' through their lateness of rent payment. This marks the end of this specific pathway, and the user is taken to the analysis page.

\begin{figure}[ht]
\begin{minipage}[b]{0.45\linewidth}
\centering
\includegraphics[width=4cm]{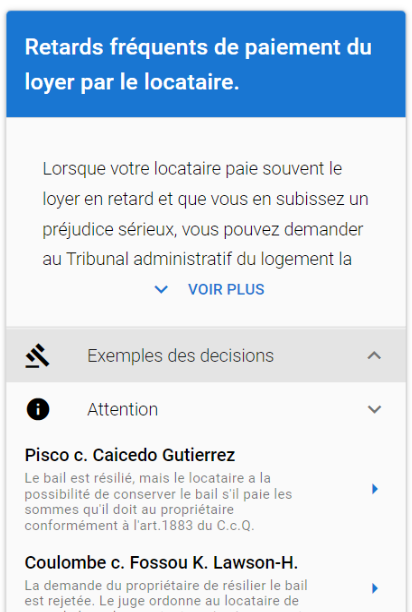}
\caption{The user is shown legal information regarding their case and previous similar cases.}
\label{fig:analysis}
\end{minipage}
\hspace{0.2cm}
\begin{minipage}[b]{0.45\linewidth}
\centering
\includegraphics[width=4cm]{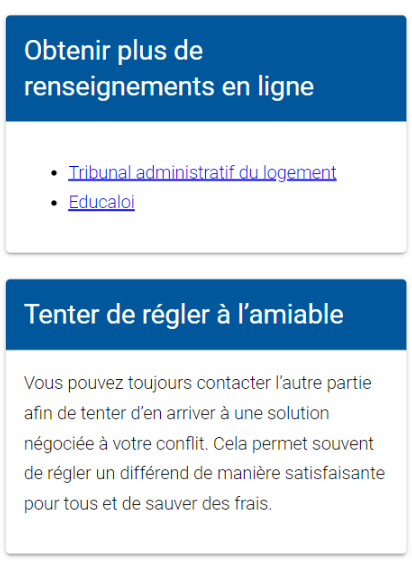}
\caption{The user is informed of next steps they can take.}
\label{fig:next_steps}
\end{minipage}
\end{figure}

The analysis screen presents the user with an analysis of their specific situation. Figure \ref{fig:analysis} shows how this information looks in the JusticeBot TAL. The user is informed that based on the answers they gave, they may have the right to terminate the lease of their tenant. Further, they are given a list of similar cases. 
The cases can be helpful in allowing the user to understand the \textit{real-world} outcomes of cases similar to theirs. The user can see that the judges terminated the lease in some cases, but in other cases decided to order the tenant to pay on the first day of the month in the future. 

Finally, the user is taken to the ``Next Steps'' page (see Figure \ref{fig:next_steps}), where they are informed of possible next steps that they can undertake to try to resolve their issue, such as speaking to a lawyer, trying to settle their situation amicably, and filing a claim at the relevant tribunal. The user is also able to review their answers, and return to a previous question to see how changing an answer may impact their rights.

We believe that the information provided by JusticeBot tools could be very useful for the individual. It may help them understand that their problem has a legal solution, which could given them a new way of resolving the problem through the legal system, e.g. by hiring a lawyer or filing a claim in court. Understanding their rights is also crucial context when seeking to find an amicable solution to their legal issues. For example, based on the information they have obtained, the user could speak to the other party and negotiate a settlement, rather than going through the arduous court process. The previous outcomes could serve as a BATNA (Best Alternative to the Negotiated Agreement) in a negotiation, compare \cite{zeleznikow2016can, dahan2020predicting}.

Now that we have seen what using a JusticeBot-based tool can look like, let us explore how such tools can be built in different legal domains, and how the system interacts with the user to present them with relevant legal information and prior cases.

\section{Proposed methodology}
\label{methodology}
This section explores the JusticeBot methodology, including how a legal expert can use it to represent legal rules and cases (section \ref{building_jb}) and how to analyze the situation of the user in light of this information (section \ref{interacting_jb}). Further, we describe the implemented tools we have built in order to encode this information and expose it to the user (section \ref{proposed_tools}).

\subsection{Encoding legal rules and cases}
\label{building_jb}
The JusticeBot methodology is inspired by the legal realism movement, and its focus on understanding the ``real rules'' that govern the reasoning of courts and can be discovered from judicial behaviour (compare \cite[p.~285]{susskind2019online}). 

Thus, the focus of the methodology is on discovering the real-world application of rules to factual situations, and exposing this information to the user in order to help them understand their situation. While legislation plays an important role in the methodology, case law may be even more important. Cases have three different uses in the methodology:
\begin{enumerate}
    \item They are used to create and validate a legal reasoning schema, that contains the steps that legal decision makers apply in making a decision.
    \item They are used to illustrate how the criteria of the law are applied, to help the user make the decision for themselves.
    \item They are used to illustrate the outcomes of previous cases that were similar to that of the user.
\end{enumerate}

\begin{figure}[t] 
    \centering
    \includegraphics[width=8.5cm]{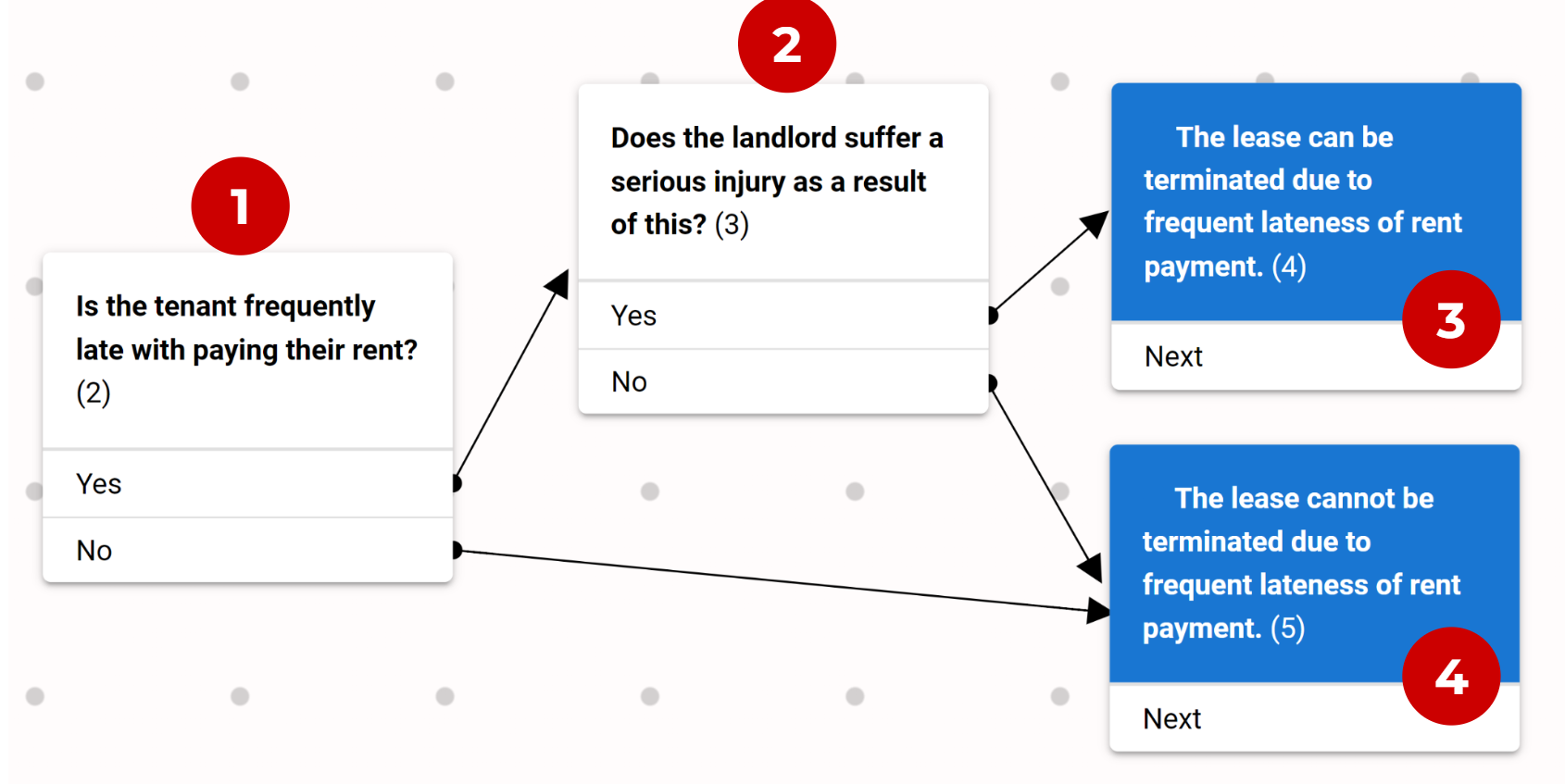}
    \caption{A legal reasoning graph for a judge deciding whether a landlord can terminate a lease due to frequent lateness of rent payment by the tenant.}
    \label{fig:reasoning_graph}
\end{figure}

We will describe each of these uses in turn below. Importantly, the steps can often be carried out simultaneously, as each single case can serve all of the three purposes. In explaining these steps, we will refer to Figure \ref{fig:reasoning_graph}, which shows the reasoning schema underlying the interaction we described in section \ref{user_flow}. This schema corresponds to part of article 1971 of the Civil Code of Quebec: 

\textit{The lessor may obtain the resiliation of the lease if the lessee is over three weeks late in paying the rent or, if he suffers serious injury as a result, where the lessee is frequently late in paying it.}

\subsubsection{Creating and validating a legal reasoning schema}
\label{sec:creating_schema}
The first step in building a JusticeBot tool is to encode the structure of the legal reasoning steps that tend to be applied to decide cases focused on a specific issue into a reasoning schema. This schema aims to capture the possible ``pathways'' that a judge can follow in reaching different conclusions. Each individual case represents a single path through this schema, as the judge applies different legal criteria in a certain order, to decide which outcome should be awarded in a case. In Figure \ref{fig:reasoning_graph}, for example, a judge might find that criterion (1) applies, and (2) does not, leading to the the outcome shown in (4).

Sometimes, these reasoning steps will stem directly from the legislation that is applied by the judge. For example, the pathway shown in figure \ref{fig:reasoning_graph} closely mimics the requirements set out in the legislation, in article 1971 of the Civil Code of Quebec. 

In other areas, the judge may apply other steps, covering multiple articles of a statute, or even adding steps that have been clarified through case law or doctrine. For example, in cases regarding an infestation of bedbugs, we found that judges apply a number of supplementary steps that are not included in the legislation to arrive at an outcome, such as whether the landlord responded quickly and diligently to the infestation, and whether the tenant collaborated with their efforts. These steps do not stem from the legislation, but would nonetheless be helpful for the user of the system, in order to understand the real-world assessment a judge would undertake. Each annotated case, of course, only shows the outcome of a single ``reasoning path'' being pursued by the judge, and thus serves as a puzzle piece to discover the overall reasoning schema.

The discovered reasoning steps can be encoded in the form of a visual representation, as can be seen in Figure \ref{fig:reasoning_graph}. The white blocks (``criterion blocks'') here correspond to a substantive legal criterion that is applied by a legal decision maker, such as ``Is the tenant frequently late with paying the rent?''. The blue blocks (``conclusion blocks'') correspond to legal conclusions that a judge can come to, such as ``The lease can be terminated due to frequent lateness of rent''.

Some criteria may build upon each other. For example, in Figure \ref{fig:reasoning_graph}, the criterion marked as (2) will only be assessed by the judge if criterion (1) is found to apply. They are thus cumulative requirements, both of which need to be fulfilled in order to arrive at a certain outcome.

A specialty of the JusticeBot methodology is that the reasoning steps carried out by the judge serve both to index previous cases, but also as an interface to the user, as we will describe below in \ref{interacting_jb}. Thus, it is important to try to keep the titles of the \textit{criterion blocks} as simple as possible. Further, each criterion block should contain a simplified explanation of the legal criterion. Collaborating with legal institutions or organisms that already provide this kind of information, and integrating their content, can be a good way of obtaining such content. Likewise, every \textit{conclusion blocks} should contain an explanation of what this legal conclusion means and why it was reached. This explanation should include an explanation of why the answers given by the user lead to a given potential outcome, in referencing relevant statutes. The content of the conclusion blocks will be shown to the user at the end of the pathway, to provide them with legal information.

In this way, the steps that judges tend to apply can be discovered and encoded in the system. When the legal expert building a JusticeBot system reads and encodes more cases, and the reasoning steps taken by a judge correspond to the schema, the correctness of the schema is validated empirically. 

\subsubsection{Illustrating the application of legal criteria}
\label{sec:illustrating_concepts}
Many of the legal criteria encoded in the schema will be open-textured terms, that have no strict definition. To help the user decide whether these criteria may apply to their situation, they are provided with so-called ``case-criterion summaries'', that show how a judge previously applied this criterion. To create these, a legal expert reads cases in conjunction with the schema, tracking which legal criterion is applied by the judge. If they find a case that could be useful for a user in understanding their factual situation, they add a summary of how the judge reasoned about a specific concept to the schema. 

This kind of summary is thus different from most types of summaries of legal cases. It aims to explain the facts that a judge considered specifically in relation to a certain legal criterion, and whether they found that criterion to apply. Thus, each single case can lead to multiple summaries, capturing how a judge reasoned in relation to different legal concepts. Likewise, the cases that are useful for this kind of summary are often ``commonplace'' cases, that illustrate how legal concepts are \textit{typically} applied, rather than ``landmark'' cases, that are extraordinary and serve as precedent for future decisions (compare \cite{stranieri1998role}).

The user can then read these summaries, and compare them to their own factual situation, in order to assess whether a judge might find the criterion to apply to their situation or not. In the JusticeBot TAL, we decided to show the user up to 5 cases where a judge did find a criterion to apply, and up to 5 where they found it not to apply.

Here are a few hypothetical example summaries of cases that could be useful for the user, relating to the criterion blocks (1) and (2) above:
\begin{itemize}
    \item The judge found a tenant not to be frequently late with paying their rent, since they had only been late 2 times in the past 5 months.
    \item The judge found a tenant to be frequently late with paying their rent, since they had been late 10 times in the past 11 months.
    \item The judge found a landlord to have suffered serious prejudice due to the frequent lateness of the tenant, since they were unable to cover the mortgage payment of the apartment for a month.
\end{itemize}

\subsubsection{Illustrating the outcomes of previous cases}
\label{sec:illustrating_outcomes}
The blue blocks shown in Figure \ref{fig:reasoning_graph} correspond to the legal conclusions that judges tend to find, based on legal criteria applying or not. For example, if the judge finds that criteria (1) and (2) in Figure \ref{fig:reasoning_graph} apply, the conclusion should be that the lease can be terminated. This information will be given to the user of the system. However, we also add summaries of the outcome of cases to the conclusion blocks, to show how previous cases that were similar were ruled upon by the judge (``case-outcome summaries'').

To represent case law outcomes, the annotator will summarize the outcome of a case. This outcome is linked to the blue conclusion presented above, representing legal conclusions. As described above, the legal expert will read the case in conjunction with the schema, following the legal reasoning pathway that the judge is on. Whenever the judge comes to a legal conclusion (e.g. ``the lease can be terminated due to frequent lateness of rent'') that directly leads to an outcome, the legal expert will create a summary of that outcome and add it to the corresponding conclusion block. Such a summary could be e.g. ``The lease was terminated.'' 

The use of text for annotating outcomes allows the annotator to capture the important aspects of the outcome of a case in a flexible manner. Here are a few examples of summaries of case outcomes:
\begin{itemize}
    \item The lease was not terminated.
    \item The judge ordered the tenant to pay their rent.
    \item The judge ordered the defendant to pay the plaintiff 500 CAD in damages.
\end{itemize}

At first glance, adding outcomes to the conclusion blocks may seem redundant, since they already contain information about potential outcomes. However, adding real life cases to the outcomes is important since judges may diverge from the mandated outcomes in certain cases (i.e. not terminating the lease and instead ordering the tenant to pay their rent on time, as per article 1973 of the Civil Code of Quebec). In order to inform the user of the system of the real-world outcomes that cases such as theirs can achieve in court, providing them with the actual outcomes is thus more helpful. Further, the case-outcome summaries are important if outcomes have a discretionary component, such as damages, which could be important for the user, and to legitimize the output of the system.



As we can see, the JusticeBot methodology is a hybrid rule-based and case-based system. Rules are used to guide the user, and represent the reasoning pathway of the judge. However, the rules are also used to index cases, that illustrate how legal criteria are applied by judges, and outcomes that were previously awarded. The next section examines how the user can interact with this system, in order to provide the details of their situation and obtain legal information. 

\subsection{Interacting with the user}
\label{interacting_jb}
In order to interact with a JusticeBot tool, the user answers questions regarding their situation and then receives information about possible outcomes and next steps. An example of a user interacting with a JusticeBot has been provided above in section \ref{user_flow}. Now, let us examine how this works on the technical level.

When the user enters a JusticeBot tool, they are taken through each criterion block in order. As we saw in section \ref{user_flow}, the first few screens serve to introduce the user to the system and generally triage their issue. 

Then, the user moves on to the specific pathways, that aim to support them in discovering their legal rights. Thus, they are taken through the specific pathway encoded as described in section \ref{sec:creating_schema}. For each criterion block, the user is shown the case-criterion summaries, whose encoding was described in \ref{sec:illustrating_concepts}. In answering the questions, the user creates a \textit{hypothesis} for how a judge may assess the different criteria in their specific situation. They can compare their situation to the previous case law, to understand how a judge may assess their factual situation. Of course, there is no guarantee that a judge will assess a new similar factual situation in the same way they did another one, which is made clear to the user. However, in general there is an  expectation that the same situation will be treated the same way in court.\footnote{Compare Lawlors concept of local and personal stare decisis - judges and courts try to exercise discretion in a consistent manner \cite{lawlor1963computers,stranieri1998role}.}

Thus, in answering the question of how the judge may apply a specific criterion to the factual situation of the user, the user poses a hypothesis of how a judge may reason about their case. This hypothesis can then be compared to actual previous cases, in order to identify similar cases that can be shown to the user to help them understand potential outcomes, should they take their situation to court.

In order to select these cases, and the legal information linked to legal conclusions, the system keeps track of whenever a user traverses a conclusion block in the schema. For each conclusion block that the user traverses, the system selects the case-outcome summaries that were added to that conclusion block, i.e. where the judge reached a specific legal conclusion. These case-outcome summaries are then shown to the user. In essence, cases are found to be similar if the judge found that certain criteria to apply, and the user expects that the same criteria should apply to their situation.

\begin{figure}[t] 
    \centering
    \includegraphics[width=8.5cm]{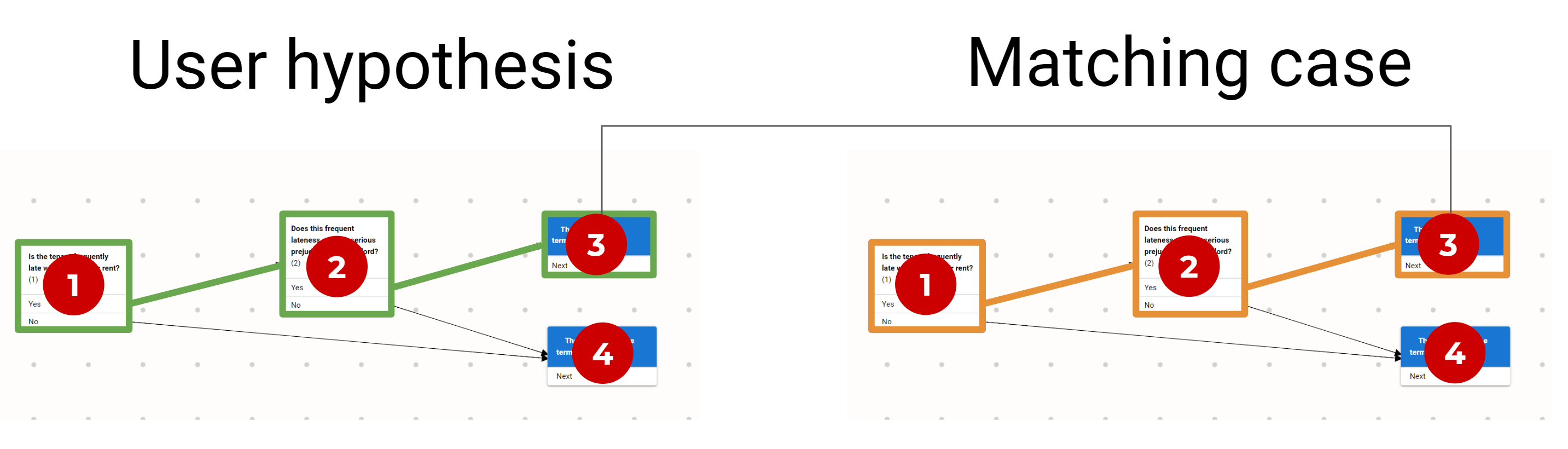}
    \caption{A relevant previous case is identified, based on the reasoning pathway taken by the judge being similar to the user hypothesis.}
    \label{fig:matching_case}
\end{figure}

Figure \ref{fig:matching_case} illustrates how these previous cases can be selected. In a previous case, the judge found criterion (1) (frequent lateness of the tenant) and criterion (2) (serious prejudice caused to the landlord) to apply. Based upon this, they logically arrived at the legal conclusion that the lease can be terminated, conclusion (3). The actual outcome of the case was thus summarized and added to the schema at block (3). A user poses the hypothesis that a judge would find criteria (1) and (2) to apply in their case as well. Thus, the system selects the previous case as a relevant case to be shown to the user at the end of their pathway. Likewise, the explanation attached to conclusion (3) is shown to the user.

As the user arrives at the end of the pathway, they are thus shown the information and previous legal cases that are relevant to their situation. The information can help them understand the possible legal consequences of their factual situation. The case-outcome summaries can help them understand the possible, real-world consequences that cases that were similar to theirs have previously lead to. This information can be important context in making a decision on how to deal with their situation.

Next, let us explore the tools we built to allow legal experts to build new JusticeBot tools, and laypeople to interact with the encoded content.

\subsection{Implemented tools}
\label{proposed_tools}
The JusticeBot methodology goes beyond a pure methodology, and also features concrete, production ready tools. The toolchain features a program that can be used to create and update JusticeBot tools, called the JusticeCreator (section \ref{JusticeCreator}), and the JusticeBot frontend (section \ref{sec:JB_frontend}), which can load the prepared data and expose it to the user. These tools have been used to build the first implemented JusticeBot tool, which will be described in section \ref{case_study}. 

\subsubsection{JusticeCreator}
\label{JusticeCreator}

\begin{figure}[t] 
    \centering
    \includegraphics[width=8.5cm]{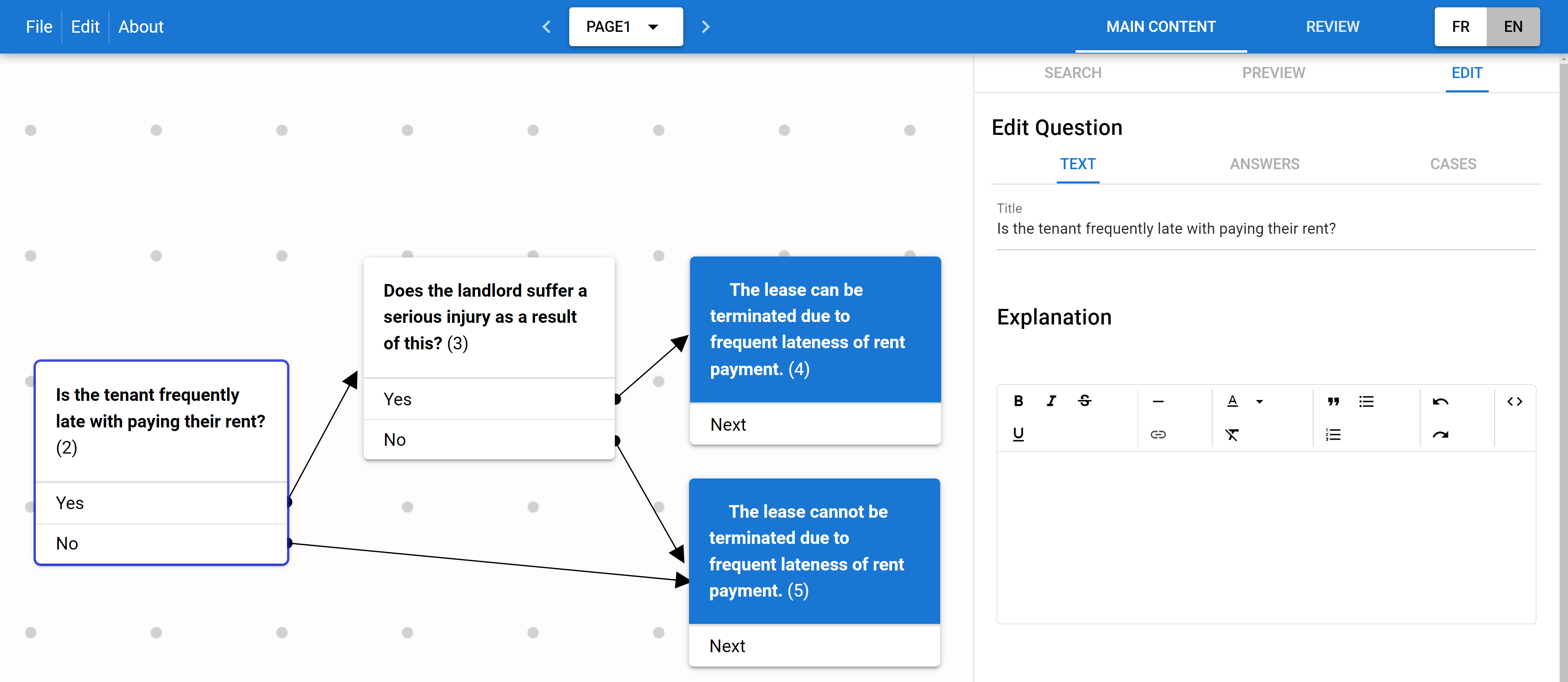}
    \caption{A screenshot of the JusticeCreator interface, which allows the creation of JusticeBot tools.}
    \label{fig:JusticeCreator}
\end{figure}

The JusticeCreator allows the creation and editing of JusticeBot tools by legal experts. Figure \ref{fig:JusticeCreator} shows the interface of the JusticeCreator. The schema view (on the left) allows the user to view the logical connections of the schema, create new blocks and logically connect blocks by dragging arrows between them. 

The panel on the right allows the editing of the selected block. The panel that is open in Figure \ref{fig:JusticeCreator} shows how the title and description of each block can be edited. The description box also allows the editing of content in a ``What You See Is What You Get'' (WYSIWYG) manner, including the formatting of the text and the adding of hyperlinks to external sources. The panel can also be used to edit the answer and add cases, as well as previewing what the system will look like for the end-user.


Identifying cases where the judge discusses a particular legal criterion can be time-intensive. The JusticeCreator includes a feature using deep learning to identify cases that could be relevant for annotation. Using a method similar to the one presented by Westermann et al in \cite{westermann2021paragraph} and \cite{westermann2020sentence}, the method encodes the title of the currently selected block using a sentence encoder, and uses a fast approximate neighborhood search method to search a corpus of encoded sentences from prior case law. Next, it surfaces the 100 cases with the most similar sentences to the title of the selected block. The legal expert annotator can read the selected cases, and choose to encode them into the schema, or decide that the cases are not relevant. This feature thus works as an ``augmented intelligence'' tool, which aims to increase the efficiency of the annotator. A more complete description and evaluation of this feature is planned for future work.

The JusticeCreator has been built with the goal of being useful for legal experts, without any technical knowledge. This is an important part of the methodology, since allowing legal experts in different legal areas to build their own legal decision support tools, without having to rely on developers, could potentially increase the number and impact of such tools, and thus increase access to justice in many areas. This ambition was validated when a team of legal experts was able to create the JusticeBot TAL, described below. After a few minutes of introduction, the individuals without technical training were able to use the system to encode the legal rules and edit the content.

Once the legal expert is happy with the schema they have created, they can export it as a JSON file. This can then be imported into the JusticeBot frontend, allowing the end-user to interact with the content.

\subsubsection{JusticeBot frontend}
\label{sec:JB_frontend}
The JusticeBot frontend is a single page application (SPA) that can be hosted online. It reads the content from the JSON-file exported by the JusticeCreator. Then, to interact with the user, the system exposes the information of the schema. Section \ref{user_flow} shows screenshots of an example interaction with the JusticeBot frontend. 

The logical steps followed by the front-end system to navigate a loaded schema are:
\begin{itemize}
    \item If the current block is a criterion block, show the title, description and associated case law to the to allow them to make a decision. Follow the selected answer to the next block.
    \item If the current block is a conclusion block, store it in a stack of information to show the user at the end of the pathway, and then immediately advance to the next connected block.
    \item If there is no more block, the pathway is finished - take the user to the analysis screen and show the stored information and associated case law to the user.
\end{itemize}

The JusticeBot frontend can be accessed either via smartphone or computer. It contains methods to collect anonymous usage statistics, and allow the user to give feedback on individual questions, their overall experience, and indicate if they have a question that is not currently covered by the tool.

Since the JusticeBot frontend takes its information from the encoded JSON-file, it is not limited to a specific domain, but can rather be used in many different domains. Currently, several additional JusticeBot tools are under development at the Cyberjustice Laboratory.

Next, let us investigate the creation process and user experience of the JusticeBot TAL, focused on landlord-tenant disputes.

\section{Casestudy: JusticeBot TAL}
\label{case_study}

\subsection{Legal area}
The first deployed version of the JusticeBot methodology focuses on landlord-tenant disputes in Quebec. Such cases are governed by the Tribunal Administratif du Logement (TAL). We were very lucky to collaborate with the TAL, which gave us access to previous decisions in bulk, and collaborated with us in the creation and launch of the tool.

Landlord-tenant disputes are a high-volume area of disputes. In 2021-2022, the TAL saw almost 64k submitted claims, and held almost 67k hearings. Further, the TAL received 930k phone calls \cite{simard_rapport_2022}. It thus seems like there is a high demand for legal information and ways to resolve conflicts in the area of landlord-tenant disputes, meaning that a JusticeBot tool could be useful for a lot of people.

\subsection{Development process \& launch}

The ``JusticeBot TAL'' focuses on providing legal information to landlords and tenants. The content of the pathway was encoded in the JusticeCreator, by consulting the law, case law and online legal information. We were very lucky to have the permission of the TAL to include legal information from their website into the JusticeBot, which provided an efficient way of integrating accurate, simplified legal information. The process was smooth, with a team of legal experts encoding the information in the JusticeCreator system after a short period of introduction. Further, the automatic method of retrieving cases proved useful in identifying suitable cases to annotate. Currently, the JusticeBot TAL contains 127 criterion blocks and 146 conclusion blocks. The tool was launched in July 2021.

\subsection{Statistics and Feedback}
Between the launch on 2021-07-20 and today, 2023-05-05, the JusticeBot TAL has been used over 20k times in total, and seen over 140k pages. The largest source of traffic was the website of the TAL, which provided almost 90\% of the users. This shows a benefit of collaborating with an institution.

The JusticeBot frontend also allows us to collect anonymized information about the questions that users answered. Over the past 60 days, 65\% of users were tenants, while 35\% of users were landlords. Table \ref{tab:tenant_stats} shows which individual pathways were picked most frequently by tenants over the past 60 days, and table \ref{tab:landlord_stats} provides the same information for landlords. Of note is the high percentage of individuals that select the ``Other'' category, which indicates that users were unable to find the appropriate pathway, or that their problem was not covered. We will return to this issue in the discussion.

\begin{table}[h!]
\centering
\caption{The percentage of tenant users who selected different issue pathways in the JusticeBot}
\label{tab:tenant_stats}
\begin{tabular}{|p{5cm}|c|}
\hline
\textbf{Pathway} & \textbf{Percentage} \\ \hline
Other & 52\% \\ \hline
My landlord wants to raise my rent & 17\% \\ \hline
My landlord wants to conduct work on my apartment & 10\% \\ \hline
I would like to leave my apartment before the end of the lease & 6\% \\ \hline
I would like to terminate my lease & 5\% \\ \hline
\end{tabular}
\end{table}

\begin{table}[h!]
\centering
\caption{The percentage of landlord users who selected different issue pathways in the JusticeBot}
\label{tab:landlord_stats}
\begin{tabular}{|p{5cm}|c|}
\hline
\textbf{Pathway} & \textbf{Percentage} \\ \hline
Other & 52\% \\ \hline
My tenant does not pay their rent & 20\% \\ \hline
My tenant is late in paying their rent & 7\% \\ \hline
My tenant has abandoned their apartment & 5\% \\ \hline
My tenant wants to leave their apartment before the end of the lease & 5\% \\ \hline
\end{tabular}
\end{table}

Users were also given the option to fill out a survey about their experience at the end of the JusticeBot pathway. In total, 35 users chose to do so. The feedback is positive. Concerning their overall experience, 63\% of the users indicated that the JusticeBot gave them the information necessary to understand their situation. 66\% of the users indicated that the JusticeBot gave them a good understanding of possible next steps regarding their situation. Finally, 89\% of the users indicated that they would recommend the JusticeBot TAL to others. This shows that even users that were not able to get an answer to their specific situation saw the usefulness of the tool.

\section{Discussion}
\label{discussion}
Now that we have explored the different components of the JusticeBot methodology, let us discuss some of the important aspects.

\subsection{Augmented intelligence}

Prior research, such as the work presented in \cite{westermann2019using}, shows that accurately predicting new, unique situations is a challenging task. Based on these insights, the JusticeBot methodology aims to augment the intelligence of the user, rather than predicting the outcome of their situation. Thus, it can be seen in the light of early research in human-computer interaction. In 1945, Bush described computers that can help users navigate the enormous amounts of information available to them \cite{bush1945we}. Likewise, Licklider in 1960 described systems that interacts with a human in a cooperative manner, with the machine complementing the capabilities of the human and vice-versa \cite{licklider1960man}.

The JusticeBot methodology aims to utilize these suggested approaches, by selecting the relevant information (in terms of legal information and previous case law) and surfacing it to the user at the right time. It acknowledges that accurately applying legal rules to factual situations is a difficult problem, and thus supports the user in carrying out this task instead of doing so autonomously. In deciding whether a certain legal criterion applies to their case, the user of the system is provided with a summary of how this criterion was applied in prior cases. This contextual information allows the user to decide whether they believe that a judge may find that a criterion applies or not to their case. Based on the answers provided by the user, the system helps them navigate the encoded legal reasoning schema.


Once the user has reached the end of a pathway, they are given a list of cases that are similar to theirs, and summaries of the outcomes. Cases are seen as being similar if the legal criteria that a judge found to apply in previous cases match what the user expects a judge would find in their case. Thus, the user can get an overview of the outcomes that were previously awarded in cases similar to theirs. This information can be very valuable to the user, as it can be interesting to use as a BATNA to negotiate a settlement, or to decide whether it is worth it to take their issue to court.

Overall, the system can thus be seen as an intelligent legal search engine. Each case is indexed by the legal reasoning schema. The user can enter a query in the form of a hypothesis how legal criteria may apply to their situation, to retrieve the outcome of cases that match this hypothesis. However, the user is at all stages informed that the system can only tell them how legal criteria were \textit{previously} applied, and the outcome that judges \textit{previously} awarded, not on how judges may act in the future. Being clear about this is important to ensure that the user understands the relevance of the information, and how to act based on it. Further, it makes it clear that the system provides legal information rather than legal advice.

\subsection{A generalizeable methodology focused on ease-of-use}
A key goal of the JusticeBot methodology is that it should be a general methodology, that could be applied in many different legal areas. Previous researchers noted the importance of designing a methodology to build e.g. legal expert systems \cite{susskind1987expert}. Following the steps described in this paper should hopefully allow the creation of many more JusticeBot tools in different legal domains. Several more tools are currently being built. Beyond legal disputes, it should be possible to apply the methodology to areas of administrative decision making, and to help people explore the legal opportunities available to them to increase their well-being, described by Susskind as ``legal health promotion'' \cite[p.~69]{susskind2019online}.

Further, to enable the simple creation of additional JusticeBot tools, the methodology was implemented in the JusticeCreator and JusticeBot frontend. The toolchain  covers all of the necessary steps from creating the pathways in an easy and intuitive manner, to deploying them to production-ready tools. Both of these components are completely reusable. The tools have further been conceptualized to be possible for legal experts to use, which can allow domain experts to create such tools, potentially without needing to rely on technically skilled individuals. For example, the encoding of legal rules in the JusticeBot methodology follows simple and deterministic rules. Reasoning about which information a user will see and why are as simple as following the schema like a flowchart, which makes it easy for the legal expert to verify that only correct information is shown to the user, without having to learn programming.

\subsection{Limitations}

Of course, there are also potential limitations with the JusticeBot approach.

First of all, it requires a relatively high amount of user efforts. In order to assess certain vague legal criteria, the user has to read cases and analyze their own situation. If they make a wrong prediction, the system may give inaccurate information at the end of the pathway.

This is a deliberate trade-off of the methodology. Predicting the outcome of a new case based on a factual situation may be quite difficult. In order to be able to build and deploy real-world systems, JusticeBot tools focus on assisting the decision-making of the user, which does require the user to perform some of the tasks. Of course, not all criteria are vague and require interpretation and case law - for example, cases may not be needed for a landlord to determine if their tenant is ``more than three weeks late with paying their rent''. Likewise, even if the user does not assess their situation correctly, they can benefit from the information on possible next steps provided by the system.

Second, as we saw, in the JusticeBot TAL a lot of users ended up choosing the ``Other'' option, indicating that their question is not covered by the system. This may be due to the user not being able to identify the pathway that is appropriate to them, or their pathway not being covered by the system. This is a limitation, but it may be less serious than it seems. First, our statistics indicate that users whose issue falls outside of the system spend an average of 30 seconds on the website. Therefore, they lose very little by trying the JusticeBot. Further, users that select the ``Other'' option are informed that the JusticeBot does not yet cover their issue, and are given the option to tell us which issue they face, which thousands of users have done. This data is an excellent source in deciding which new issue to add to the system. Thus, JusticeBot tools can be progressively enhanced to cover more and more situations.

Third, the system may be difficult to apply in areas where there are complex legal reasoning steps. If cases tend to involve diverging reasoning steps for the same situation in different cases, it may be difficult to fully capture a reasoning schema that works for most cases. This is a limitation. However, the JusticeBot methodology is initially intended to be used in areas of issues that laypeople tend to face, such as areas of high-volume, low-intensity decisions or administrative decisions. Here, it is more likely that a consistent reasoning schema can be identified and used to capture cases.

The reader may also ask whether the system can be employed in both common law and civil law jurisdictions. The first version of the JusticeBot was created in the area of landlord-tenant disputes, which follows the civil law rules in the Civil Code of Quebec. The civil law is often said to rely more heavily on statutes and rules. Especially in first instance courts, it would seem likely that the courts often apply a common reasoning schema to cases, stemming from the law. The interpretation of the judge may lie in whether a certain set of facts fulfills a given legal criterion, which is perfectly suited to the way the JusticeBot models legal areas, rather than interpreting the structure of the law.

The common law, on the other hand, seems to more heavily rely on analogical reasoning with previous cases (compare e.g. \cite{ashley1991reasoning}), which may be harder to model with the JusticeBot methodology, since they may not have a strict schema that judge tend to follow. That said, it is very possible that the JusticeBot approach could still be useful in some areas of high-volume, low-intensity disputes, where the legal tests are perhaps more consistent, and argument with cases plays a lesser role.

\subsection{Access to Justice?}

Now that we have discussed some of the aspects of the JusticeBot methodology, let us analyze whether it can be seen to fulfill its ultimate goal of supporting access to justice.

As we discussed in section \ref{intro}, laypeople often have trouble resolving their legal issues, which can have significant negative impacts on the individuals and society. Laypeople may not be aware that their issue has a legal solution, or how to enforce their rights. People that end up taking their case to court often struggle with the process, including knowing which forms to fill out, or which facts they need to prove.

The JusticeBot allows layperson users to select a situation or goal that they wish to achieve, and explore the legal rules linked to this situation or goal. The  tool guides the user through the different legal criteria, asking them to provide the specifics of their situation, and then provides them with legal information, outcomes of previous cases, and possible next steps. 

This information could help the user better understand the laws and rules that apply to them. For one, this could help them arrange their affairs to comply with the rules, which could increase the rule of law, and the feeling of belonging of individuals to society. 

Likewise, individuals could use the information to settle their case with the other party, by negotiating an amicable solution in light of the previous outcomes and rights. Or, they could decide to hire a lawyers and/or go to court. As an increased number of legal issues can be solved, societal welfare and trust in legal institutions may increase.

The numbers indicated by the JusticeBot survey are promising. Over 60\% of the users indicated that the information they were given gave them an understanding of their situation and possible next steps. Any users gaining an increase in understanding of their legal situation represents a potential improvement of the status quo, and we are excited that a majority indicated to have obtained such an increased understanding through the use of the system. Further, almost 90\% of survey respondents indicated that they would recommend the tool to others, which is a significant endorsement.

Of course, it is important to remain humble. Access to justice is a societal, inter-disciplinary issue, which is likely to require significant investment and attention to fully solve. However, hopefully, the presentation of this methodology can lay at the basis of many legal decision support tools with the potential to inform laypeople and give them better ways resolve their legal problems and thereby increase overall welfare in society.

\section{Future Work}
\label{future_work}
There are many ways to build and expand upon the work presented in this paper. More JusticeBot tools could be built in other areas, such as consumer issues, employment issues, debt issues, immigration questions, various government license procedures, and obtaining social aid. In all of these areas, helping the public understand the applicable rules could help many people. JusticeBot tools may also target other users beyond laypeople, such as lawyers, judges, legal aid clinic workers or government employees.

There could be many ways to improve and expand the core platform as well. This could include making interaction with user easier, e.g. by integrating machine learning methods and large language models to link the user situation to the correct pathway. It could also include ways to make the creation of pathways more efficient, e.g. by automatically classifying previous cases, or automatically creating summaries of cases (compare \cite{salaun2022conditional}).

Further, the JusticeBot methodology and toolset could be used to output documents instead of information, or be integrated in an Online Dispute Resolution platform to support settlement or guide the user towards the correct procedure.

\section{Conclusion}
\label{conclusion}

We introduced the JusticeBot methodology, which aims to increase access to justice. The methodology can be used to create legal decision support tools, that ask a user questions to understand their situation, and then provides them with legal information, summaries of case law, and possible next steps. We described the details of the methodology and its implementation and reported initial results from the first implemented JusticeBot version.

\begin{acks}
We would like to thank the Cyberjustice Laboratory at Université de Montréal, the LexUM Chair on Legal Information and the Autonomy through Cyberjustice Technologies (ACT) project for their support of this research. Further, we would like to thank the Tribunal Administratif du Logement and the Ministère de l'Économie et de l'Innovation for their collaboration on the project. We are further grateful to all of the members of the JusticeBot team at the Cyberjustice laboratory.
\end{acks}

\bibliographystyle{ACM-Reference-Format}
\bibliography{sample-sigconf}


\end{document}